\documentclass[largeformat]{interact}

\usepackage{epstopdf}
\usepackage[caption=false]{subfig}

\usepackage[doublespacing]{setspace}
\usepackage[numbers,sort&compress]{natbib}
\bibpunct[, ]{[}{]}{,}{n}{,}{,}
\renewcommand\bibfont{\fontsize{10}{12}\selectfont}

\theoremstyle{plain}

\theoremstyle{definition}

\theoremstyle{remark}

\begin{document}

\title{The Role of the Greco-Roman Practice as a Progenitor of the Armenian and Eastern Roman Ornamental Art}

\author{
\name{Mehmet Erbudak\textsuperscript{a} and Selim Onat\textsuperscript{b}\thanks{CONTACT: M. Erbudak: erbudak@phys.ethz.ch -- S. Onat: selim.onat@free-now.com\\The analysis and dataset presented in this manuscript can be downloaded from the following repository https://github.com/selimonat/ornament\_symmgroups}}
\affil{\textsuperscript{a}Department of Physics, ETHZ, CH-8093 Zurich, Switzerland and
Physics Department, Bo\u{g}azi\c{c}i University, Istanbul, Turkey; \textsuperscript{b}Free Now Tech, Data Science, D-22767 Hamburg, Germany}
}

\maketitle

\begin{abstract}
We investigate two-dimensional, periodic ornaments of the Late Hellenistic (some centuries before the Common Era, the Classical Period) and Early Roman (Common Era) classical periods found at different locations in Asia Minor in Turkey and classify them into mathematical wallpaper groups based on their symmetry properties. The source material comes from Terrace Houses in Ephesus, Izmir, from  Zeugma, now in the Zeugma Museum, Gaziantep, and from the recently released bathing pool in Antiochia ad Cragum near Gazipa\c{s}a, Antalya. Using  the artifacts  we first determine the occurrence of each symmetry group. Then we compare this distribution with those of the medieval cultures of the Middle East, namely  the Armenian,  Byzantine,  Arab and Seljuk Turks, calculating in pairs the Euclidean distances of the wallpaper distributions. The subsequent multi-dimensional scaling and hierarchical cluster analysis of the results  confirm  that the Armenian and Byzantine artworks are strongly inspired by the classical masterpieces, as is the Seljuk creation by the Arabs.

\end{abstract}

\begin{keywords}
Ornaments; mosaics; wallpaper group; Byzantine; Armenian; Hellenistic; Roman; Greco-Roman; Arab; multi-dimensional scaling; hierarchical cluster analysis;
\end{keywords}


\section{Introduction}

Ornaments, especially regularly tessellated patterns, are a collection of intelligent signs created by men from a cultural group. They are means of expression and basic tools of human abilities, which are passed on to others. This information is characteristic of the group of craftsmen or architects who  express their state of mind, which has developed and accumulated over time. Therefore, the ornaments can be examined  to identify an ethnic or cultural group. The classical school uses the decorative elements, the  harmony of colors and shapes as  key criteria for the  classification of ornaments. This approach has the advantage of preserving the beautiful details of the artwork. Owen Jones \cite {Jones}  classified the ornaments  in his encyclopedic work according to  the ethnic or cultural groups in which they were created.

Symmetry brings out some fascinating  qualities in  works of art. These are two-dimensional, periodic properties of ornaments, which can be described by  their point-group and translational symmetries. The novel achievement consists in mathematically establishing 17 {\em wallpaper groups\/} within group theory. This kind of categorization made it possible to extract invariant features of artworks that seemed to be very different on one level of perception \cite {Doris}.

The idea of characterizing planar patterns by wallpapers can be applied to objects created by different  civilizations. The goal of such efforts is motivated by the question whether the use of symmetries of  patterns and their distribution across the 17 wallpaper groups is characteristic of the culture that created them \cite{WashCrowe}. If so, this knowledge could be used to determine the origins of the craftsmen and the cultural interactions. For example, one hypothesis is that the more  cultural relations and exchanges between civilizations occur, the more similar the artworks created by these civilizations could be, and therefore the more similar the  wallpaper categories are.

In a previous work we  determined the distribution of wallpaper groups based on Armenian, East Roman (Byzantine), Arab and Seljuk ornaments. A statistical similarity analysis shows a close resemblance between the symmetry properties of the Arabic artwork and those of the Rum Seljuks \cite{MS1}. We have interpreted the result as a strong interaction between these cultural groups. The dedication to their commitment helped the Seljuks embrace the artistic skills of the Arabs and culminate the graphic design into two-dimensional masterpieces that go  beyond its simple form  of small repetitive units \cite {Mehmet1}.

Our investigation of several cultures in the Middle East \cite{MS1} was limited to the Middle Ages, i.e.,  after the iconoclasm in the Eastern Roman Empire,  with an open  question about the origin of several artistic achievements. In this paper we investigate the possibility of the ancient Hellenistic culture flourishing before 400 BC and the later Roman influence around the turn of the millennium as the source of inspiration for the later Armenian and Eastern Roman art. Asia Minor is a treasure trove of ancient Hellenistic-Roman (Greco-Roman) cultural sites from the Aegean coast in the west to Mesopotamia in the southeast of modern Turkey. However, there are only a few sites with a good number of well-preserved ornamental pieces to ensure a certain statistical certainty. 

For our studies, we use mosaics found in three locations. First, we consider those mosaics that were discovered in the {\em Terrace Houses\/} of Ephesus, Selçuk, Turkey. The area was first cultivated around 500 BC, i.e., after the Archaic Period. Later during the Classical Period (around 200 BC) the  Houses were improved and finally inhabited with its full facilities until the 7th century during the Roman Empire \cite{Tabanli}. The next archaeological site is  {\em Zeugma\/}, which is today  in the {\em Gaziantep Zeugma Museum\/}, because the whole ancient  site is flooded by the construction of the Birecik Dam on the Euphrates River in 2000. The Hellenistic twin cities of  Seleucia Apamea and Zeugma were built on opposite banks of the river by Seleucos Nicator, one of the four generals of Alexander the Great;  together they are renamed  {\em Zeugma\/},  in the Greek language the {\em bridge\/} of the boats that crossed the river at that point \cite{Zeug1}. The third ancient site is the Hellenistic city {\em Antiochia ad Cragum\/} with the largest Roman mosaics, founded around 170 BC near Gazipa\c{s}a in southern Turkey \cite{Dodd}.

The next Section is devoted to the details of the ornamental symmetries found in these three sites. In  Section 3 we use these  objects of the Classical Period to provide  general information about the symmetries that were found  before the Middle Ages. The comparison of symmetry distributions of the Greco-Roman artwork  with those of the Middle Ages is the topic of the next Section. Here we use the statistical idea of {\em Euclidean distance\/} \cite{MS1} to create the basis for the multi-dimensional scaling (MDS) \cite {Kruskal} and {\em hierarchical cluster analysis\/} (HCA)  \cite{Jain} algorithms. Finally, we show how close the symmetries between the classical Anatolian cultures and  the East Roman and  Armenian peoples are. This connection is analogous to the influence of the Arab/Islam creations on the Seljuks of Rum.


\section{Materials}

The classical and medieval mosaics that decorate a monument usually show illustrative scenes from the Old Testament, Greek mythology or some typical scenes of the local culture they represent. Usually the areas of the illustrations are separated with decorative friezes, one-dimensional periodic ornaments. More rare are the two-dimensional ornaments. We search for such ornaments and classify them according to their symmetries into 17 different wallpaper groups \cite {Doris}. 

We have noticed the close connection between the Seljuk and the Arabic decorative arts in terms of their symmetries \cite{MS1}. Here we look for the source of inspiration of  medieval Armenian and Byzantine art. Most of the Armenian  architectural work is associated with stonemasonry craft. Armen Kyurkchyan has a rich archive of such ornaments, which he  collected for more than 30 years in  historical Armenia  \cite{Kurk}. We  have applied the classification scheme to these Armenian ornaments and identified their symmetry properties \cite{Erbudak1}.

The Eastern Roman Empire had its capital in Constantinople with a large extension from Arabic countries to Sicily. We  have searched the church of Hagia Sofia in  Istanbul for ornamental mosaics and could not find any, except for some wall paintings, maybe frescoes, and four golden mosaics in fourfold symmetry on small areas that decorate  the emperors'  robes. We are not sure about the origin of the wall paintings; they could have been created during the numerous repair works on the huge building. On the contrary, the number of mosaics seem to have decreased from 37 to 13 after the large repair work of Gaspare Fossati  $1847-1849$ \cite {Mango}. Thomas Whittenmore, an expert on Byzantine culture, writes "... we are reluctantly compelled to believe that it was shaken down by the devastating earthquake which destroyed so many mosaic paintings in Hagia Sophia on the 10th of July 1894" \cite {Whittenmore}.

Since there is no usable material with symmetry in Hagia Sofia in Istanbul,  we  applied this classification method to the San Marco Cathedral in Venice \cite {Erbudak2}   in order to have a standard for  Eastern Roman practice. This cathedral was built after the Church of the Holy Apostles in Constantinople \cite {Dumb} and is a standing example of  Eastern Roman church architecture.

\vspace {3mm}
\begin{figure*}[h]
   \centering
\includegraphics[width=8.2cm]{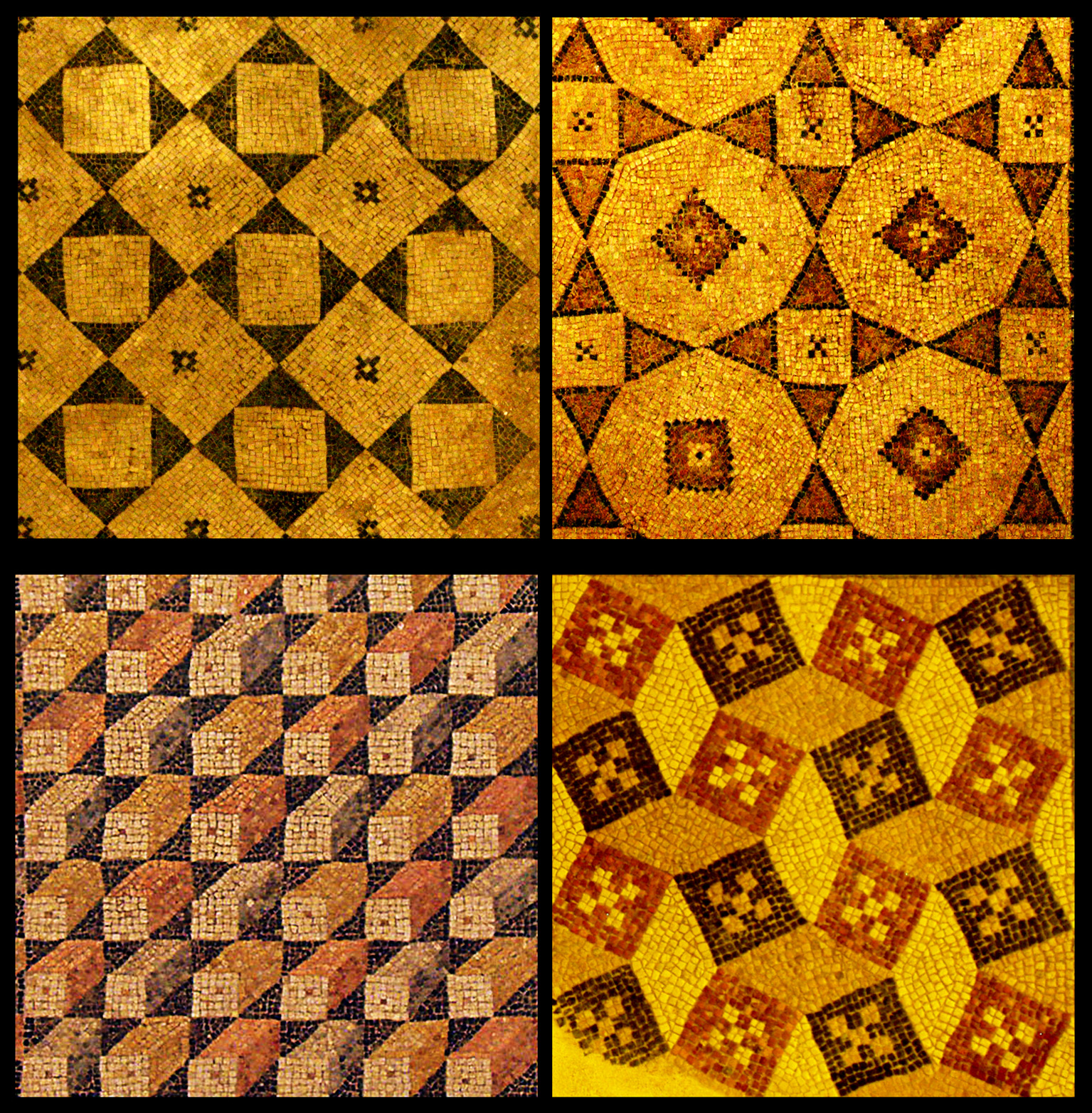}
    \caption{Floor mosaics with two-dimensional, periodic ornaments found (upper row) in the Terrace Houses in Ephesus and (lower row) in the Zeugma Museum.}
    \label{mosaic}
\end{figure*}

We conducted a detailed statistical analysis on the ornaments created by the cultures of the medieval Middle East and attributed the similarity of the symmetry distributions to the common religious belief of the peoples involved. In  search of further common inspiration, we present here the classical Greco-Roman practices found in Asia Minor.

Figure \ref{mosaic} presents  two examples  from Ephesus and two from Zeugma. They are two-dimensional periodic ornaments, which  are well preserved: The upper two from the Terrace Houses in Ephesus both have the symmetry of the group {\em p4mm\/}, while the examples from Zeugma have {\em p1\/} (bottom left) and {\em p4gm\/} (bottom right) symmetries. Mosaics of  Antiochia ad Cragum are shown in an arial survey of the archaeological site \cite{Cragum}. For the Greco-Roman  civilization we found 21 two-dimensional periodic ornaments in the Terrace Houses in Ephesus, 32 such ornaments in the Zeugma Museum, and 13 in  Antiochia ad Cragum. Neglecting multiple occurrences, we looked at  61 Greco-Roman ornamental artifacts. 


\section{Results}

\begin{figure*}[t!]
   \centering
\includegraphics[width=7.0cm]{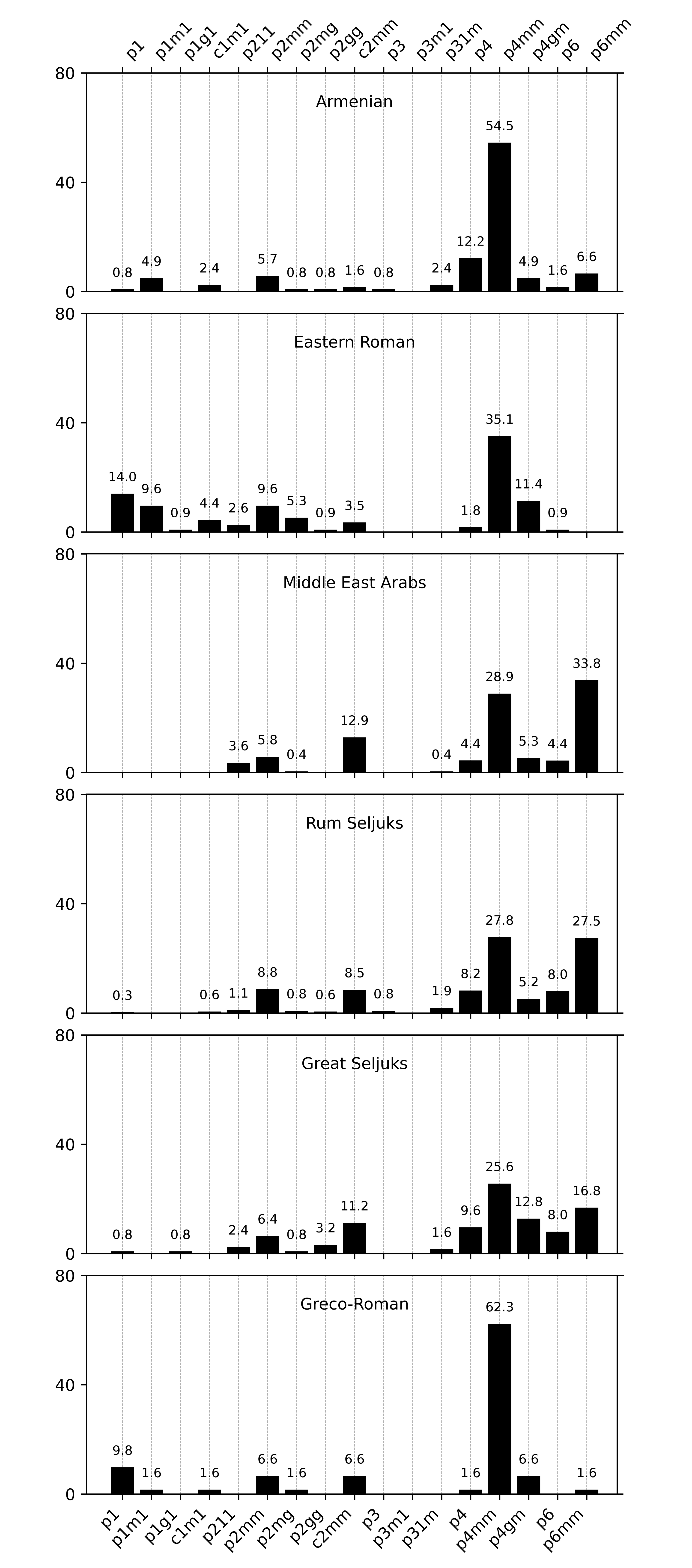}
    \caption{Occurrence of the individual symmetry groups (in percentage) found in the planar ornaments of six civilizations of the Middle East.}
    \label{histogram}
\end{figure*}

The frequencies of the symmetry groups found in the ornaments created by the Middle Eastern Arabs, Great Seljuks, Rum Seljuks, East Romans, and Armenians were presented in our earlier communication \cite {MS1} and are shown here in Figure \ref {histogram} together with the corresponding data for the classical cultures designated as Greco-Roman (lower panel). The Greco-Roman distribution is characterized by a strong peak at the {\em p4mm} group and a small number of observations for the {\em p6mm} group.

We investigated the similarity relationship between the newly introduced Greco-Roman cultural group and the 5 cultures  we   analyzed previously. As we already applied a normalization to convert the counts into percentages, we did not consider the Pearson correlation as a similarity metric (such as $1-r$, {\em r\/}: Pearson correlation). This would have introduced a second normalization by the standard deviations of each group and could possibly distort the data. Instead, however, we used the Euclidean distance $d$ as in our earlier work. This was calculated across the 17 symmetry groups between two cultures $X$ and $Y$ as follows:
$$d_{XY} =  {\sqrt {\sum_{i=1}^{17} (x_i - y_i )^2 \/}} $$ 
Euclidean distances are bounded by zero at the lower-end, and values closer to zero indicate low dissimilarity (or high similarity) between two groups of observations, $X$ and $Y$. In this analysis, small Euclidean distances between two cultures result from a high similarity between the distributions of the symmetry groups and are therefore interpreted as an indication of shared, common cultural art practices in the production of ornaments. On the other hand, high Euclidean distances arise when the distributions are less similar and can be interpreted as divergent practices in the production of ornaments.

\begin{figure*}[h]
   \centering
\includegraphics[width=16 cm]{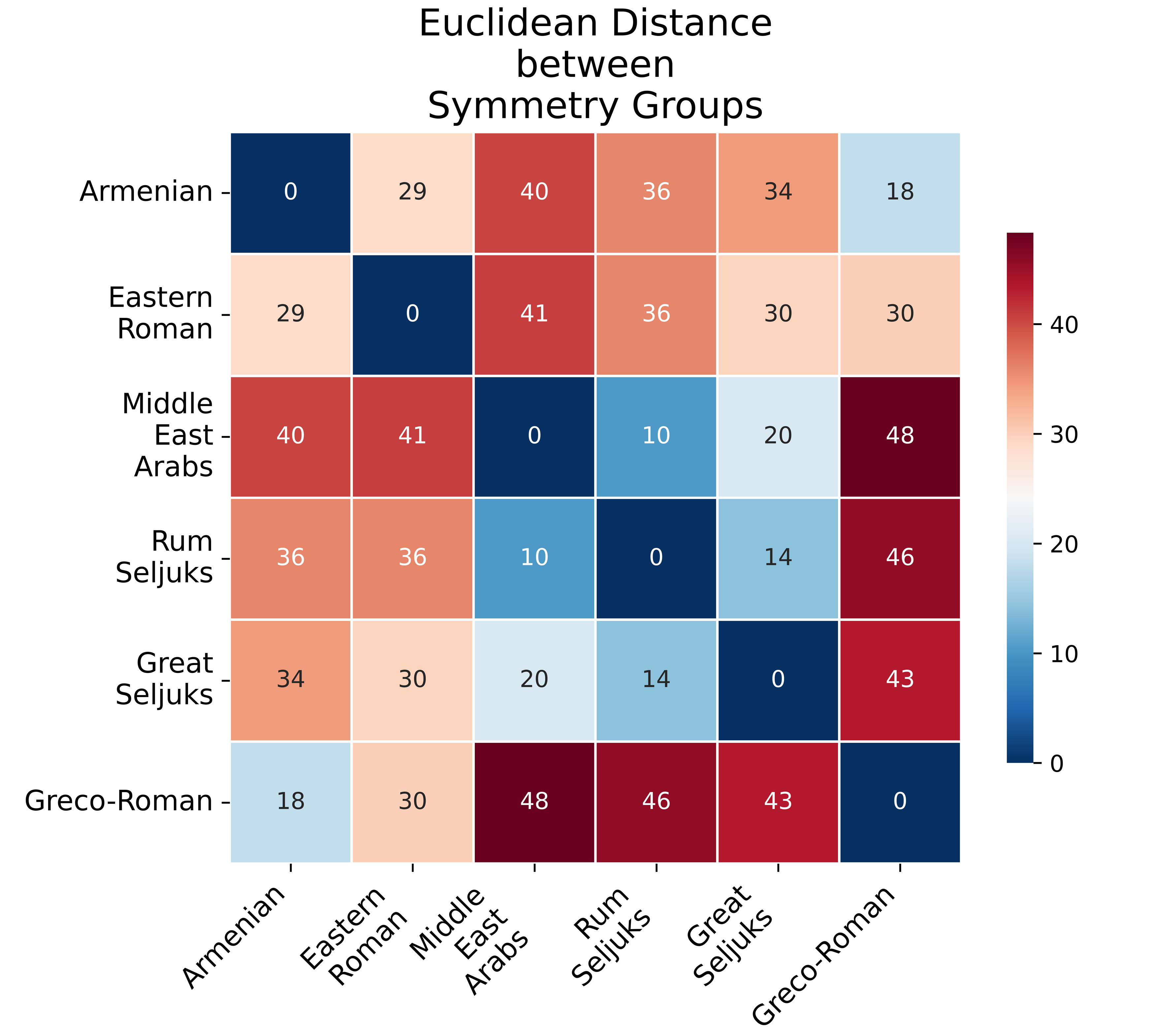}
    \caption{Color-coded Euclidean distances $d$ between all pairs among 6 cultural groups. Smaller values indicate greater similarity. The diagonal measures the self-similarity between the cultural groups, resulting in a dissimilarity of 0. Note that the matrix is symmetrical.}
    \label{distance}
\end{figure*}

The calculated Euclidean distances between all culture pairs are given in Figure~\ref{distance} in a matrix form. We find that the strongest similarity (Euclidean distance 10.3) exists between the Middle East Arabs and Rum Seljuks. The closest cultures to the newly-analyzed Greco-Roman group are the  Armenians (Euclidean distance 18.2) and the Eastern Romans (Euclidean distance 30.0). The smallest distances between the groups are shown in Table~\ref{high}. In this table, we also present the Pearson correlation values in the third column, since these are generally more intuitive.

We first used the MDS method \cite {Kruskal} to better understand similarity relationships between the 6 cultures and especially to place the Greeco-Roman group in relation to 5 cultures that we had previously investigated.

MDS aims to create a summarizing representation of a dissimilarity matrix. To achieve this, MDS finds the best positions for each group that is present in the dissimilarity matrix, and does so  in such a way that distances between these points reflect their dissimilarity values as close as possible. In general, these points are embedded in a coordinate system with at most 3 dimensions (to allow visualization), which excludes a perfect solution. Instead, the positions are found by minimizing a stress term using an iterative optimization \cite {Kruskal}.

\vspace{4 mm}
\begin {table}[h]
\begin{center}
\begin{tabular}{|c|c|c|} 
\hline

Cultural Groups & $d$ & $r$ \\ 
\hline\hline
Arab -- Rum Seljuk&10.3&0.97\\
\hline
Great Seljuk -- Rum Seljuk  &14.2 &0.92\\ 
\hline
Greco-Roman -- Armenian  &18.2 &0.96\\ 
\hline
Great Seljuk -- Arab  &20.3 &0.88\\ 
\hline
Eastern Roman -- Armenian   &28.8 &0.85\\ 
\hline
Eastern Roman -- Great Seljuk  &29.5 &0.59\\ 
\hline
Greco-Roman -- Eastern Roman  &30.0 &0.93\\ 
\hline
\end{tabular}\\
\end{center}
\caption{Values of Euclidean distances $d \leq 30.0$ of  ornamental symmetries for 6 cultural groups in descending order of similarity. This is the list of groups with strongest cultural interactions. To simplify the interpretation,  we also show the Pearson correlation $r$ between the corresponding cultures, which varies between -1 and 1.}
\label{high}
\end {table}

\begin{figure*}[h]
   \centering
\includegraphics[width=9.6cm]{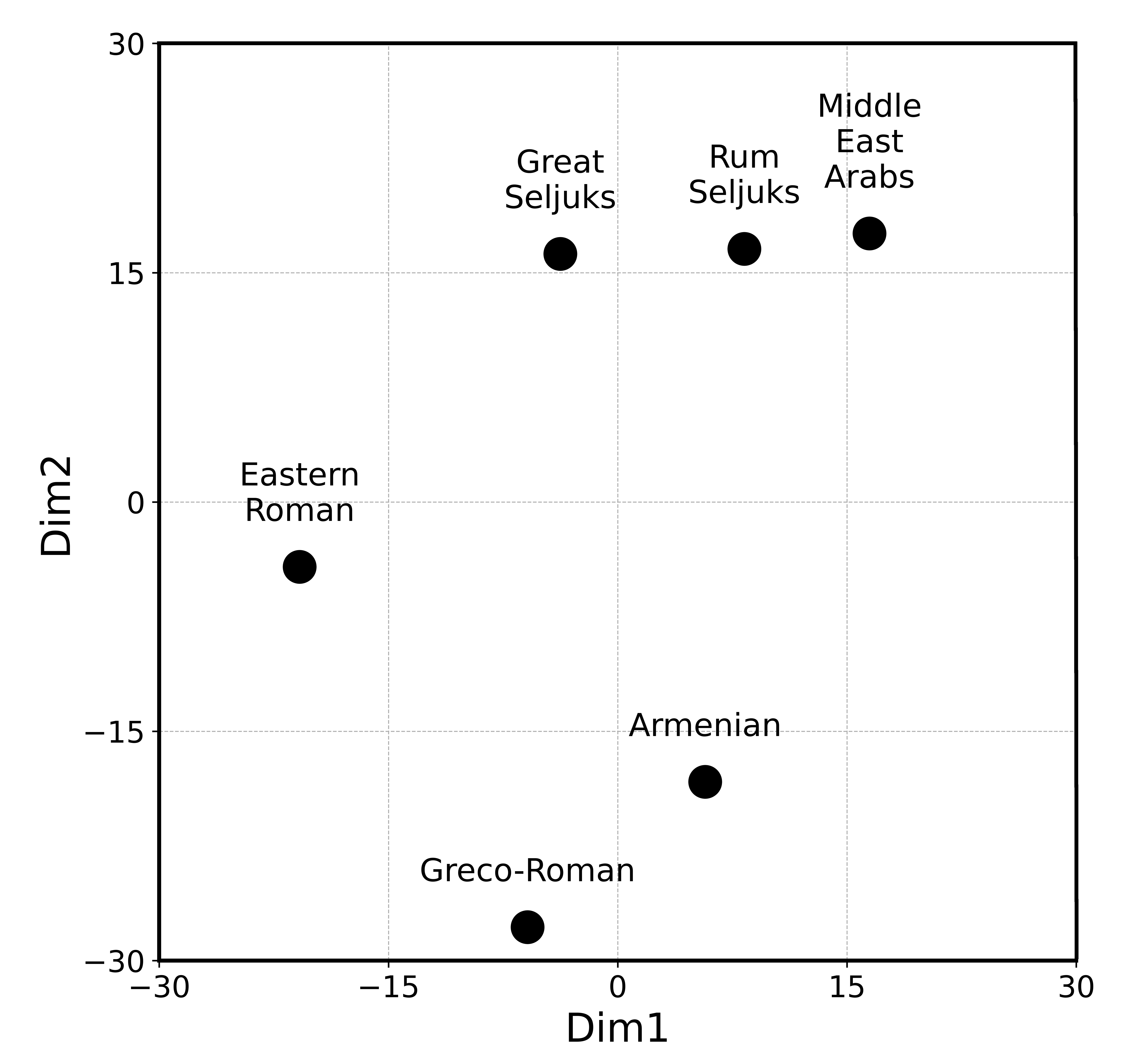}
    \caption {Two-dimensional MDS plot of the Euclidean distance between six cultural groups. The axes with arbitrary units  are a product of the MDS calculation.}
    \label{MDS}
\end{figure*}

In this work we used two-dimensional metric-MDS using {\em scikit-learn\/} implementation in Python (ver: 0.23.2). Figure~\ref{MDS} is the result of a two-dimensional MDS calculation using the data given in Figure~\ref{histogram}. Metric-MDS iteratively minimizes  a term called {\em metric stress} which is the squared sums of the differences between the positions of a pair of points and the corresponding dissimilarity values. Due to its iterative nature, the final solution depends on the initial starting points. We used 50 different runs with random starting points and here we present the one that yielded the smallest stress value, namely 0.057 (Kruskall's normalized stress value). We also performed a one-dimensional MDS analysis and observed a normalized stress value of 0.22, which is usually considered an indication of poor fits.

This analysis shows that the ornamental artifacts created by  cultural groups with Islamic religious background are all placed close to each other around Dim2 = 15,  indicating their increased similarity. On the MDS representation, Middle East Arab and Greco-Roman cultures are the most widely separated; they are the two extremes. The Armenian and, to a lesser extent, the Eastern-Roman work of art are closer to the Eastern-Roman creation. This observation gives rise to the conjecture that Armenian and Eastern Roman artworks form a continuity with the earlier cultural artifacts created during the Hellenistic and Roman periods in Anatolia several centuries ago.

\begin{figure*}[h]
   \centering
\includegraphics[width=8.2cm]{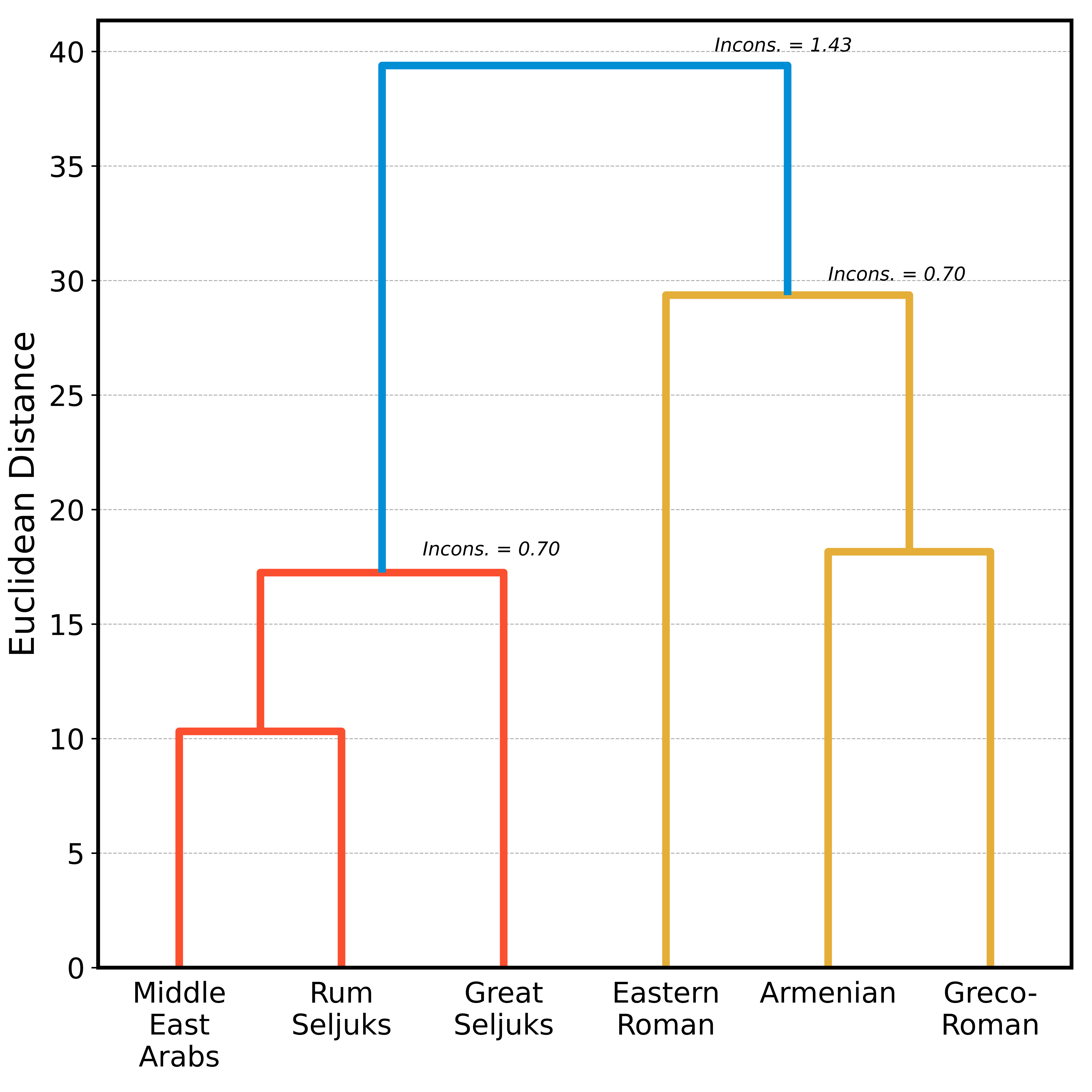}
    \caption {A dendogram based on the dissimilarities between the symmetry groups of medieval cultures.
    The Euclidean distance is given in arbitrary units.  It is obvious how two discrete red and yellow clusters  are formed.}
    \label{dendro}
\end{figure*}

To elucidate this conjecture, we have used the hierarchical cluster analysis \cite{Jain} and investigated whether the similarity relationships  shown above are compatible with the presence of clusters. Based on the similarity relationships of objects, HCA generates an upside-down tree-like structure, called {\em dendrogram\/}. The dendrogram is built iteratively by linking similar objects together (with horizontal lines, called {\em nodes\/}) and aggregating the dissimilarity values of the linked objects. This linkage and aggregation process is repeated until there are no more objects to be linked. In the resulting dendrogram, shown in Figure \ref{dendro}, the value of the y-axis where nodes are located shows the dissimilarity between linked elements, and the vertical lines are called  branches. For the type of hierarchical clustering one can use the {\em maximum\/} (or {\em complete\/}), the {\em average\/} and the {\em minimum linkage method\/} \cite{Jain}.

The dendrogram illustrated in Figure \ref{dendro} results from the dissimilarity matrix shown in Figure~\ref{distance}.  In this work we used the average of linked objects as the aggregation operation. The {\em cophenetic correlation} evaluates how well the distances between the objects in a dendrogram correlate with the original dissimilarities between same objects and is thus a  measure of quality for the linking operation. We observed a cophenetic correlation of 0.91, indicating that our dendrogram was a true representation of our original dissimilarities. The use of the popular {\em complete linkage method\/} resulted in a cophenetic correlation of 0.87. The two-dimensional MDS plot presented in Figure~\ref{MDS} shows that the distances between cultures sharing the Islamic  background are smaller compared to other 3 cultures and form a cluster. The dendrogram plot in Figure \ref{dendro} clearly shows that the Greco-Roman culture together with the Armenian and Eastern Roman cultures are  grouped in a separate branch from the three Middle Eastern cultures with Islamic background (Seljuks and Arabs) and confirms our prediction about this additional cluster.

Dendrograms allow to investigate the presence of clusters in a data set. Intuitively, the presence of strong clusters can be detected by the presence of long parent-branches (vertical lines) connected to small child-branches. Therefore, the mixture of long and small branches can be an indication of strong clustering in a data set. The {\em inconsistency coefficient} measures the heterogeneity of the dissimilarities on a given link level. It compares the dissimilarity at the link level with the average dissimilarity of child links, normalized by their standard deviation. Large inconsistency values indicate that the dissimilarity observed at a given linkage level is greater than what could be expected only from the  child branches and can therefore  be used as evidence for the presence of naturally occurring clusters. 

At the top level, we observed an inconsistency coefficient of 1.43 (shown in Figure \ref{dendro}). This value was higher than the inconsistency observed for links at a hierarchical level below, namely 0.7. We interpret the observed inconsistency coefficient at the top level as an indication for the presence of moderate clustering in our data set. The Arabs and the two Seljuks show a grouping around 20 on the y-axis, while the Greco-Roman culture forms a cluster with the Armenian and Eastern Roman cultures around 30 signalling the formation of a loose group.

 
\section{Conclusions}

The symmetry analysis of the ornaments created by earlier cultures is interpreted as characteristic of these cultural groups. The results of the MDS indicate two  different groupings. One group includes the Islamic Arabs, while the Seljuk Turks closely followed this group in the production of their ornaments. The second group is formed by  the classical Greco-Roman creation with Armenian and  Eastern Roman (Byzantine) cultures in close proximity.

The Greco-Roman group as a source of inspiration is actually no surprise. The cultural development in  Asia Minor, especially around the {\em Fertile Crescent\/}, and simultaneously around the Aegean Sea, began several millennia before the present. We read of the Assyrians, Babylonians, Hittites and Sumerians as the first known cultures before the Hellenistic colonization of  Asia Minor. Urartu was about a millennium before Christ. Later, the Persian invasion and the subsequent Macedonian conquest under the  leadership of Alexander the Great acted like a sponge and opened  the interior of Asia Minor to Hellenistic settlements and influence.  Roman annexation took place  around the turn of the millennium, but Hellenistic culture remained dominant. The Eastern Roman Empire (Byzantium) and the Armenian Kingdom were founded long before Seljuk Turks immigrated from Central Asia after the Battle of Manzikert in 1071. In the period that  followed,  Anatolia  gradually experienced a transition from predominantly Christian and Greek speaking to predominantly Muslim and Turkish speaking territory. We can therefore claim that Byzantine or Armenian art practices developed in countries that were strongly influenced by Hellenistic culture. Our work merely supports this cultural historical sequence.

\vspace{8 mm}
\noindent
This research received no external funding.



\begin{thebibliography}{1}

\bibitem {Jones} 
Owen Jones, {\it The Grammar of Ornament\/}, Dorling Kindersley, London, 2001.\\https://doi.org/10.1515/9781400882717.

\bibitem{Doris}
Doris Schattschneider, {\em The Plane Symmetry Groups: Their Recognition and Notation\/}, Am. Math. Mon. {\bf 85},  $439-450$ (1978). https://doi.org/10.1080/00029890.1978.11994612.

\bibitem {WashCrowe}
Dorothy K. Washburn and Donald W. Crowe, {\em Symmetries of Culture: Theory and Practice of Plane Pattern Analysis\/}, University of Washington Press, Washington (1991) pp. 229. \\https://doi.org/10.1080/00029890.1991.12000774.

\bibitem{MS1}
Mehmet Erbudak and Selim Onat, {\em Similarity in Symmetry Groups of Ornaments as a Measure for Cultural Interactions in Medieval Times\/}, https://doi.org/10.20944/preprints202008.0031.v2.

\bibitem{Mehmet1}
Mehmet Erbudak, {\em Mathematical Classification of Rum Seljuk Ornaments\/},  Symmetry: Culture and Science {\bf 2\/}, $177-198$ (2020). https://doi.org/10.26830/symmetry\_2020\_2\_177.

\bibitem{Tabanli}
Didem Tabanlı, {\em Roma Dönemi Mozaiklerinin Efes Örneğinde İncelenmesi\/}, Masters Thesis, Ege University, Institute of Fine Arts, İzmir, 2007.  http://hdl.handle.net/20.500.12397/9794

\bibitem{Zeug1}
https://whc.unesco.org/en/tentativelists/5726/

\bibitem{Dodd}
Emlyn Dodd, {\em Late Roman viticulture in Rough Cilicia: an unusual wine-press at Antiochia ad Cragum\/}, J. Roman Archaeology {\bf 33},  $467-482$ (2020). https://doi.org/10.1017/S1047759420001129

\bibitem {Kruskal}
Joseph B. Kruskal,  {\em Multidimensional scaling by optimising goodness of fit to a nonmetric hypothesis\/},  Psychometrika {\bf 29}, $1-27$ (1964). https://doi.org/10.1007/BF02289565

\bibitem{Jain}
Anil K. Jain and Richard C. Dubes,  {\em Algorithms for Clustering Data\/}, Prentice Hall, Englewood Cliffs (1988) Section 3.3.6,  pp. $121-122$; Charles T. Zahn, {\em Graph-theoretical methods for detecting and describing Gestalt clusters\/}, IEEE Transactions on Computers (1971) C-20(1):68-86; Daniel M\"ullner, {\em Modern hierarchical, agglomerative clustering algorithms\/}, arXiv:1109.2378v1; Ziv Bar-Joseph, David K. Gifford and Tommi S. Jaakkola, {\em Fast optimal leaf ordering for hierarchical clustering\/}, Bioinformatics, 2001. https://doi.org/10.1093/bioinformatics/17.suppl\_1.S22

\bibitem{Kurk}
Armen Kyurkchyan (Author), Hrair Hawk Khatcherian (Photographer), {\em Armenian Ornamental Art\/}, Craftology, Yerevan, 2010.

\bibitem {Erbudak1} 
Mehmet Erbudak and Armen Kyurkchyan,  {\it Armenian, Byzantine and Islamic  Ornaments -- Influences Among Neighbours\/}; Kyurkchyan: Yerevan, Armenia, 2019. https://doi.org/10.3929/ethz-b-000394011

\bibitem {Mango}
Cyril Mango, {\em The Lost Mosaics of St. Sophia, Constantinople\/}, Actes du XXIIe Congr\`es International des \'Etudes Byzantines, Pt. 3, pp. $227-234$ (1964).

\bibitem{Whittenmore}
Thomas Whittenmore, {\em The Mosaics of Hagia Sophia at Istanbul\/}, Oxford University Press, Oxford (1942).

\bibitem {Erbudak2} 
Mehmet Erbudak,  {\em Symmetry analysis of the floor ornaments of the San Marco cathedral in Venice\/}, Heliyon  {\bf 5},  e01320 (2019).  https://doi.org/10.1016/j.heliyon.2019.e01320

\bibitem {Dumb}
{\em The Holy Apostles, Visualizing a Lost Monument\/}, Dumbarton Oaks Publications, Washington DC (2015); https://www.doaks.org/holy-apostles/; {\em Constantine of Rhodes, On Constantinople and the Church of the Holy Apostles\/}, Liz James, ed., Ashgate, Farnham (2012).

\bibitem{Cragum}
Ben Kreimer, {\em 3D Modeling the Antiochia ad Cragum - Archaeological Research Project in Turkey.\/} http://benkreimer.com/aerial-video-photography/antiochia-ad-cragum-aerial-survey





\end{thebibliography}
\end {document}